\documentclass{PoS}
\usepackage{amsmath} 
\usepackage{booktabs}

\def\fm {\mathop{\hbox{fm}}}
\def\MeV {\mathop{\hbox{MeV}}}

\newcommand{\beq}{\begin{equation}}
\newcommand{\eeq}{\end{equation}}
\newcommand{\beqa}{\begin{eqnarray}}
\newcommand{\eeqa}{\end{eqnarray}}

\title{Neutron electric polarizability}

\ShortTitle{Neutron electric polarizability}

\author{\speaker{Andrei Alexandru}\\
       Physics Department, The George Washington University, Washington, DC 20052, USA\\
       E-mail: \email{aalexan@gwu.edu}}

\author{Frank X. Lee\\
       Physics Department, The George Washington University, Washington, DC 20052, USA\\
        E-mail: \email{fxlee@gwu.edu}}
       
\abstract{We use the background field method to extract the ``connected'' piece of the neutron 
electric polarizability. We present results for quenched 
simulations using both clover and Wilson fermions and discuss our experience in extracting 
the mass shifts and the challenges we encountered when we lowered the quark mass. For the
neutron we find that as the pion mass is lowered below $500\MeV$, the polarizability starts rising
in agreement with predictions from chiral perturbation theory. For our lowest pion mass,
$m_\pi=320\MeV$, we find that $\alpha_n = 3.8(1.3)\times 10^{-4}\fm^3$, which is still only
one third of the experimental value. We also
present results for the neutral pion; we find that its polarizability turns negative for pion
masses smaller than $500\MeV$ which is puzzling.}

\FullConference{The XXVII International Symposium on Lattice Field Theory\\
                July 26-31, 2009\\
                Peking University, Beijing, China}

\begin{document}

\section{Introduction}

Hadron polarizabilities measure the ability of the electromagnetic field to deform the hadrons. At the lowest order, the
effect of the electromagnetic field on the hadrons is parametrized by the effective Hamiltonian:
\beq
H_{em} = -\vec{p}\cdot\vec{E} - \vec{\mu}\cdot \vec{B} - \frac{1}{2}(\alpha E^2 + \beta E^2) + \ldots,
\label{eq1}
\eeq
where $\vec{p}$ and $\vec{\mu}$ are the electric and magnetic dipole moments and $\alpha$ and $\beta$ are the 
electric and magnetic polarizabilities. Due to the time reversal symmetry of the strong forces, the electric dipole moment
of the hadrons is zero and the electric field effects are quadratic in the strength of the electric field. The electric
polarizability, $\alpha$, measures the induced electric dipole and it is the most important parameter needed to describe
the interaction of weak electric fields with the hadrons.

Experimentally, the electric polarizability is measured in Compton scattering off nuclear targets -- the experiments
allow access only to a combination of electric and magnetic polarizabilities and further modeling is required
to disentangle these parameters. Extracting the polarizabilities for the neutron is especially difficult since there are no
free neutron targets; the best experimental numbers come from neutron scattering off lead \cite{Schmiedmayer:1991zz} 
and deuteron \cite{Kossert:2002ws}. The experimentally determined value for the electric polarizability is 
$\alpha_n = 12.5(2.5)\times 10^{-4} \fm^3$.

\begin{figure}[t]
\begin{center}
\includegraphics[width=8cm]{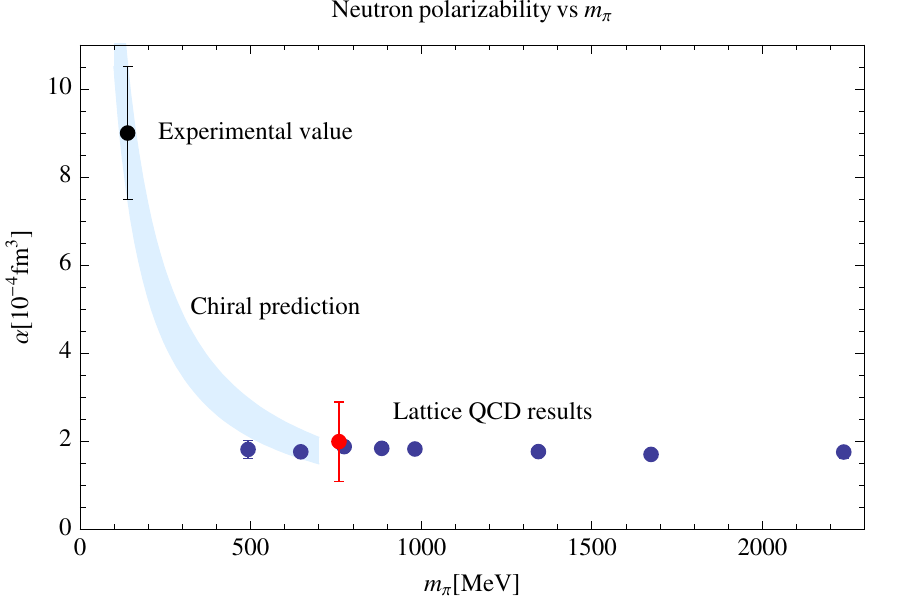} 
\caption{Neutron electric polarizability: the red and blue points are lattice QCD results
\cite{Engelhardt:2007ub,Alexandru:2008sj} and the black point is the experimental result
\cite{Beane:2004ra}. The chiral prediction is a $1/m_\pi$ fit that is forced to go through
the experimental value and asymptote to the lattice QCD values.
 \label{fig1}}
\end{center}
\end{figure}

Lattice QCD calculations of electromagnetic polarizabilities have a long history: the first study was published 20 years
ago \cite{Fiebig:1988en}. This study employed the background field method and used quenched staggered fermions and rather 
heavy quark masses; however, the results looked promising. Further studies employing quenched wilson and clover 
fermions found again good agreement with the experimental value \cite{Christensen:2004ca}. Recently, the interest in the problem was
renewed. A number of groups are trying to determine the polarizability in the chiral limit 
\cite{Engelhardt:2007ub, Detmold:2008xk, Detmold:2009dx, Alexandru:2008sj} and new issues have surfaced. 
Some problems with the 
method used to introduced the background field were resolved \cite{Alexandru:2008sj, Shintani:2006xr} and the new studies
produce much smaller values for the neutron polarizability. The new values are no longer in
good agreement with the experimental value as we can see from Fig.~\ref{fig1}; however, chiral perturbation theory predicts
that the electric polarizability diverges in the chiral limit and it is possible that the disagreement is only due to the relatively
heavy quark masses used in the current lattice QCD simulations. 

Our ultimate goal is to compute the electric polarizability of the neutron and other hadrons in the physical limit and to compare
the lattice predictions with the experimental results. The purpose of the current study is to determine the methodology to 
compute electric polarizability; in particular we focus on the fitting methods used to extract the electric mass shifts and
the range of quark masses needed to be able to perform the chiral extrapolation. To clarify this last point:
since simulations using the physical quark masses are not yet feasible, chiral extrapolations will need to
be used. We have to determine how small the quark masses need to be to observe the chiral behavior. 
From our previous study~\cite{Alexandru:2008sj}, we know the neutron polarizability is rather flat for
pion masses larger than $500\MeV$; in the current study we use a pion mass down to $320\MeV$.
Chiral  perturbation theory predicts that the divergent behavior will also appear in quenched simulations \cite{Detmold:2006vu} and
since the purpose of the current study is exploratory in nature, we use quenched ensembles.

The plan of the paper is the following: we start by briefly reviewing the background field method; we then
present the fitting method we used to extract the mass shifts and compare it with the method used in our
previous study. We use the neutral pion as an example because the signal is cleanest in the pseudoscalar
channel. We present our results for the neutron polarizability and conclude by discussing our plans for the
future.

\section{Background field method}

In this section we describe the method we use to compute the electric polarizabilities:
the background field method. The electromagnetic field is introduced via minimal
coupling, i.e. the covariant derivative in the presence of the electromagnetic
potential $A_\mu$ is
\beq
D_\mu = \partial_\mu -i g G_\mu - i q A_\mu, 
\eeq
where $G_\mu$ is the color field. On the lattice, this amounts to modifying the links:
\beq
U_\mu \rightarrow U_\mu e^{-i a q A_\mu}.
\eeq
The polarizability is determined by measuring the change in the mass of a hadron
when we turn the background field on. From Eq.~\ref{eq1} it can be inferred that the
shift in the energy is negative for a positive electric polarizability. This is still true on the
lattice if the electric field is introduced via a real factor~\cite{Alexandru:2008sj,Shintani:2006xr}.
On the lattice it is convenient to work with imaginary factors; in this case the relationship 
between the energy shift and the polarizability needs to be adjusted. For example, if
we introduce an electric field in the $x$ direction using $U_x \rightarrow U_x e^{-i a q E t}$
then
\beq
\Delta m = +\frac{1}{2}\alpha E^2.
\eeq

\begin{figure}[t]
\begin{center}
\includegraphics[width=8cm,trim= 12cm 0cm 0cm 8cm,clip=true]{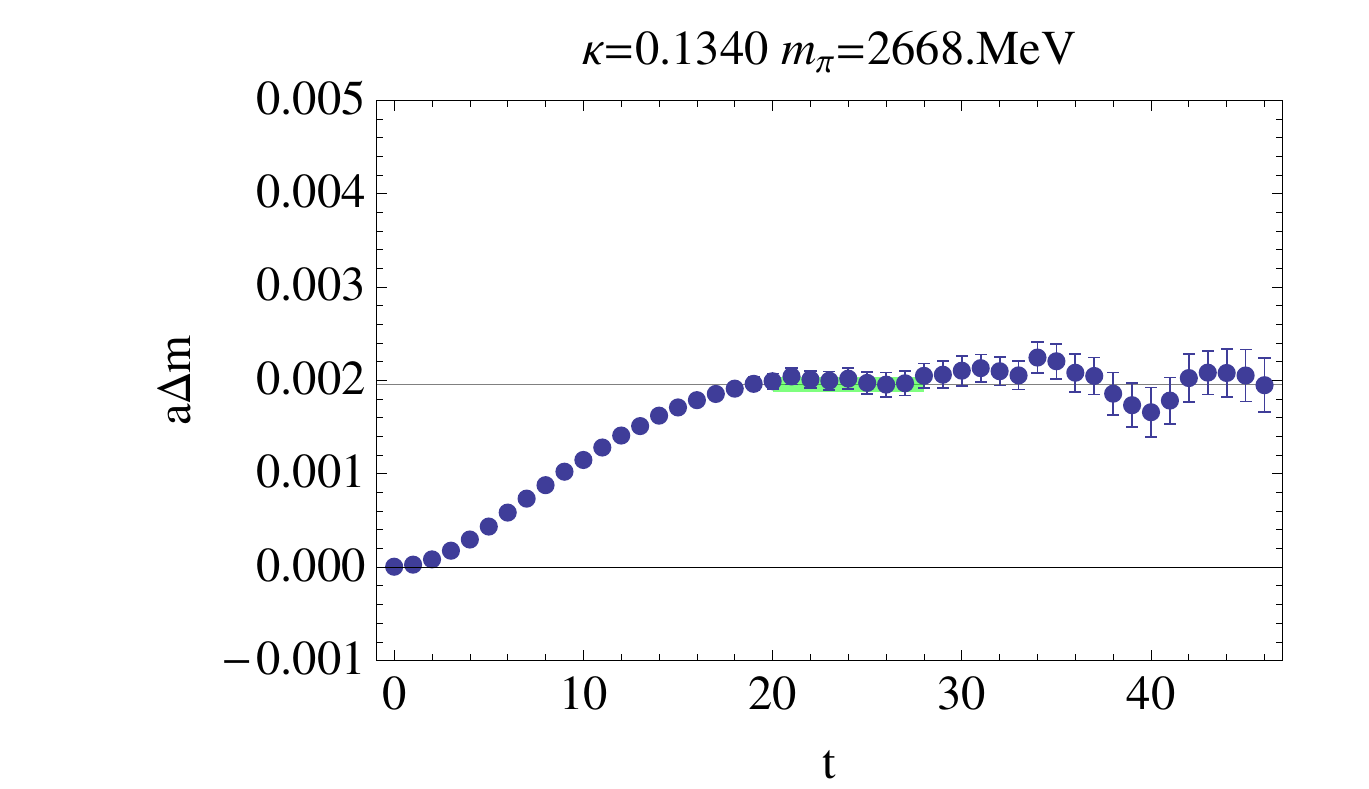} % requires the graphicx package
\caption{The effective mass shift for a heavy pion; 
this is based on the ratio defined in Eq.~{\protect \ref{eq4}}.
\label{fig2}}
\end{center}
\end{figure}

The effect of the electric field on the hadron masses is very weak and the mass shifts
we have to measure are minute. In fact, the mass shifts are usually much smaller than the
stochastic error in the determination of the hadron masses. To extract such tiny shifts we rely
on the correlation between the hadron propagators measured on the same color gauge background.
Denote with $G_0(t)$ the hadron propagator of interest and with $G_E(t)$ the same propagator
measured with the electric field turned on. The error of each of these propagators is larger than
their difference, but they are strongly correlated. To measure the mass shift we look at their ratio
which for large times is expected to be dominated by the lowest energy in the channel:
\beq
R(t)=
\frac{G_E(t)}{G_0(t)}
\xrightarrow
{\mbox{\tiny $t\rightarrow\infty$}} 
\frac{A' e^{-m' t}}{A e^{-mt}} \propto 
e^{-\Delta m t}.
\label{eq4}
\eeq
We can define an ``effective mass" based on this ratio, $a \Delta m_{\rm eff}\equiv-\ln R(t+1)/R(t)$, and see whether 
this quantity exhibits a plateau. In Fig.~\ref{fig2} we show such an effective mass plot for
the neutral pion; for this heavy quark case, we get a very good plateau
and we can use a single exponential fit in the range $t\in[20,40]$ to extract the mass shift. 
At lower quark masses the propagators become noisier
and more sophisticated techniques are needed to extract the mass shift.

\section{Extracting the mass shift}

\begin{figure}[t]
\begin{center}
\includegraphics[width=10cm]{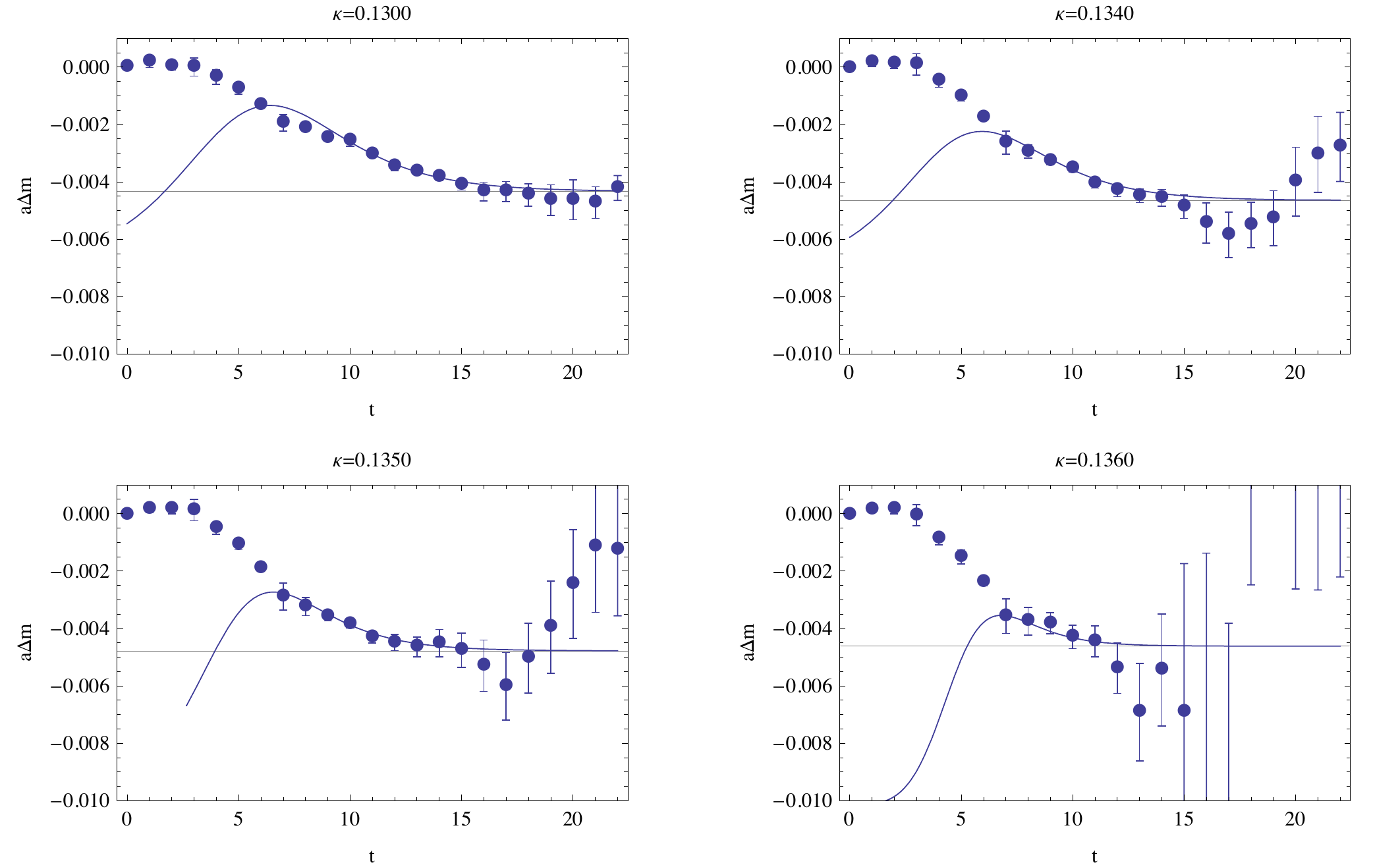} % requires the graphicx package
\caption{The effective mass shift for neutron on $24^4$ lattices; 
the solid line represents the results of the fit.
\label{fig3}}
\end{center}
\end{figure}

As we emphasized in the last section, the mass shift induced by the electric field is tiny and we need
special methods to determine it. Ideally, we would be able to generate propagators that produced 
good plateaus for the effective mass shift. In our previous study~\cite{Alexandru:2008sj} we used
$24^4$ lattices with $a=0.093\fm$. In Fig.~\ref{fig3} we present the effective mass plots for the
mass shift of the neutron (note that the mass shift are negative because in that study we introduced
the electric field using a real factor). We cannot detect any plateau in these figures -- the
reason for it is two fold: for heavier quark masses it looks like the lattice is too short to form a plateau and
for lighter masses the signal gets noisy before we can detect a plateau.

To solve this problem, we
used a two exponential fit to extract the mass shift. Since we are fitting the ration of two propagators,
the two exponential fit involves 6 independent parameters:
\beq
\frac{G_E(t)}{G_0(t)}
\rightarrow
\frac{W_1' e^{-m_1' t}+W_2' e^{-m_2' t}}{W_1 e^{-m_1 t}+W_2 e^{-m_2 t}} =
\frac{A_1 e^{-\delta m_1 t} + A_2 
e^{-\delta m_2}}{1+A_3 e^{-\delta m_3 t}}.
\eeq
Since 6 parameter fits were too unstable, we extracted $A_3$ and $\delta m_3$ from a two
exponential fit for $G_0(t)$ and used these values as priors for the 6 parameter fit. The results of
these fits are plotted with solid lines in Fig.~\ref{fig3}. 

\begin{figure}[t]
\begin{center}
\includegraphics[width=11cm]{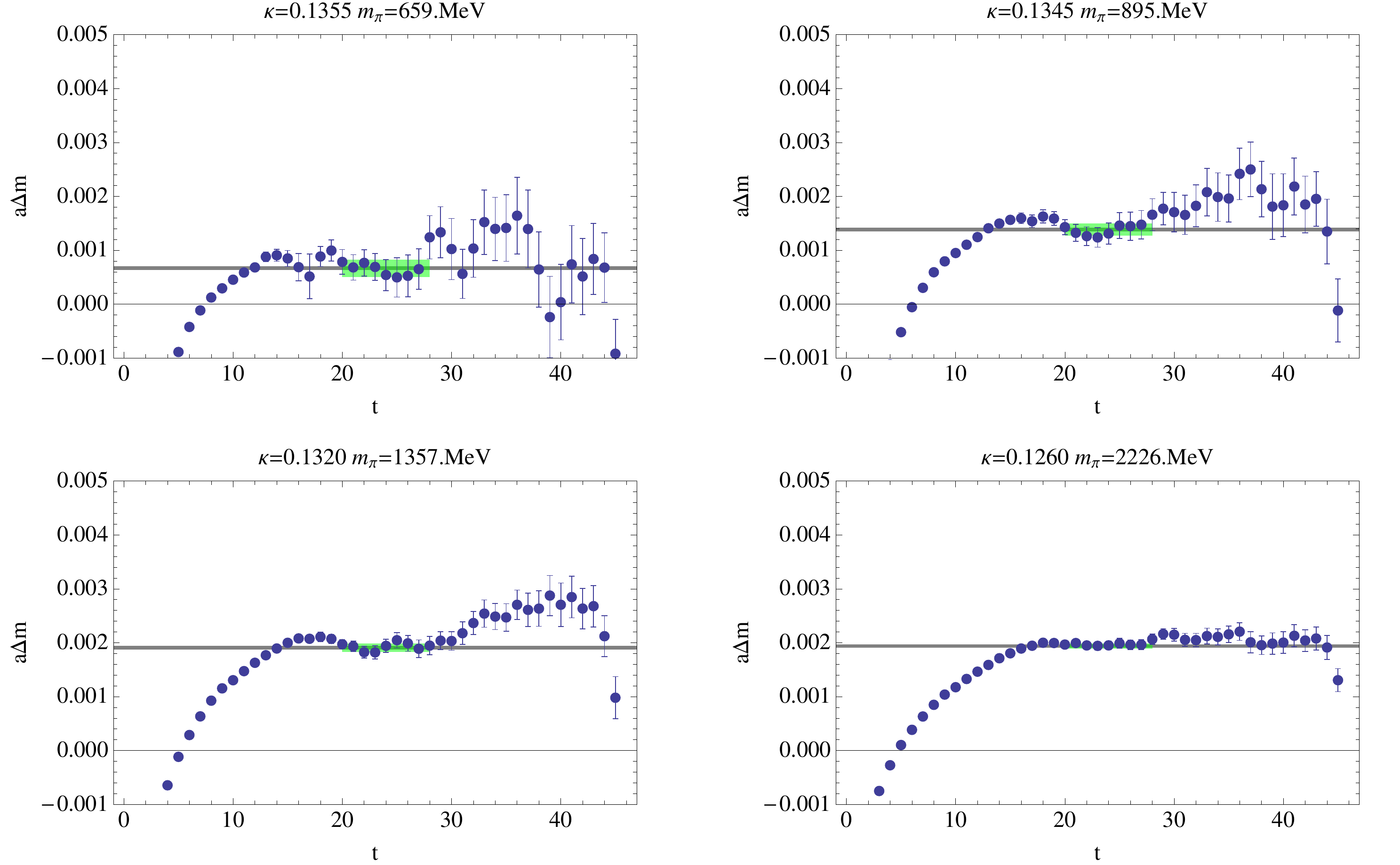} % requires the graphicx package
\caption{The effective mass shift for neutral pion compared with the result of the fit. The solid
lines indicate the results of the fit and the green boxes' height indicate the error estimate;
the width of the green boxes is used to indicate the range of the fit. The fit agrees with the
plateaus in the effective mass plots.
\label{fig4}}
\end{center}
\end{figure}

For the new set of runs we used lattices of size $24^3\times 48$, so that we can observe the plateau,
at least for heavy quark masses. We also used a different method to extract the mass shift. The main
reason we decided to use a different method is that the double exponential method used above reuses
the $G_0(t)$ data and this forces us to use time-consuming jackknife or bootstrap when estimating the errors. 
Moreover, the method proved cumbersome to use when we decided to scan the 
fitting range; these scans are needed to understand how robust the results are. The new method
we used is faster and more stable -- it allows us to scan the ranges efficiently and also it produces
error estimates that are trustworthy. The method is simply based on a correlated fit that uses both
$G_0(t)$ and $G_E(t)$ data simultaneously; the difference vector is defined to be:
\begin{eqnarray}
\nonumber
\delta_i &\equiv& f(t_i)-G_0(t_i),\quad\quad\quad {\rm for}\quad i=1,\dots,n \\
\delta_i &\equiv& \bar{f}(t_{i-n}) - G_E(t_{i-n}), \quad {\rm for}\quad i=n+1,\dots,2n
\end{eqnarray}
where $f(t) = A \exp(-m t)$ and $\bar{f}(t) = A' \exp(-(m+\Delta m)t)$ and $n$ is the number of time slices 
used in the fit. The fitting function then minimizes $\chi^2 = \frac{1}{2}\delta^T C^{-1} \delta$, where $C$ 
is the $2n\times 2n$ correlation matrix. The two $n\times n$ diagonal blocks of this matrix are the usual correlation
matrices for $G_0(t_i)$ and $G_E(t_i)$ and the off-diagonal blocks represent the cross-correlations (which
are strong since these propagators are evaluated on the same configurations). The fit depends on 4
parameters and $\Delta m$ is the mass shift we want -- this also allows us to get the error bars for the
mass shift directly from the $\chi^2$ fitting analysis, a procedure that is both fast and with a solid theoretical 
foundation. In Fig.~\ref{fig4} we show the results of fitting the pion propagators using this method
and compare it with the plateaus seen in the effective mass plot; the agreement is encouraging. This is the
method that we employ to compute the mass shift for the results presented in this paper.

\section{Simulation parameters}

All our simulations are run on quenched ensembles, generated using Wilson gauge action with $\beta=6.0$;
this corresponds to a lattice spacing of $a=0.093\fm$. We had three sets of runs, and the relevant information
is presented in Table~\ref{tab1}.

% Requires the booktabs if the memoir class is not being used
\begin{table}[htbp]
   \centering
   %\topcaption{Table captions are better up top} % requires the topcapt package
   \begin{tabular}{@{} ccccc @{}} % Column formatting, @{} suppresses leading/trailing space
      \toprule
      Ensemble    & fermion type &number of configurations & pion mass range ($\MeV$)& exceptional\\
      \midrule
      ${\cal E}_1$  & clover & 1000  & 500 -- 2200 & 34\\
      ${\cal E}_2$  & Wilson & 300 & 580 -- 2700  &  0\\
      ${\cal E}_3$  & Wilson & 200   & 320 -- 750     &  4\\
      \bottomrule
   \end{tabular}
   \caption{The parameters for the three ensembles used in this study.}
   \label{tab1}
\end{table}

The boundary conditions are important for the background field method. It turns out that a constant field
is not easy to introduce on a lattice with periodic boundary conditions without creating a discontinuity. In this
study we use Dirichlet boundary conditions in both time direction and the direction of the electric
field. 

\begin{figure}[t]
\begin{center}
\includegraphics[width=11cm]{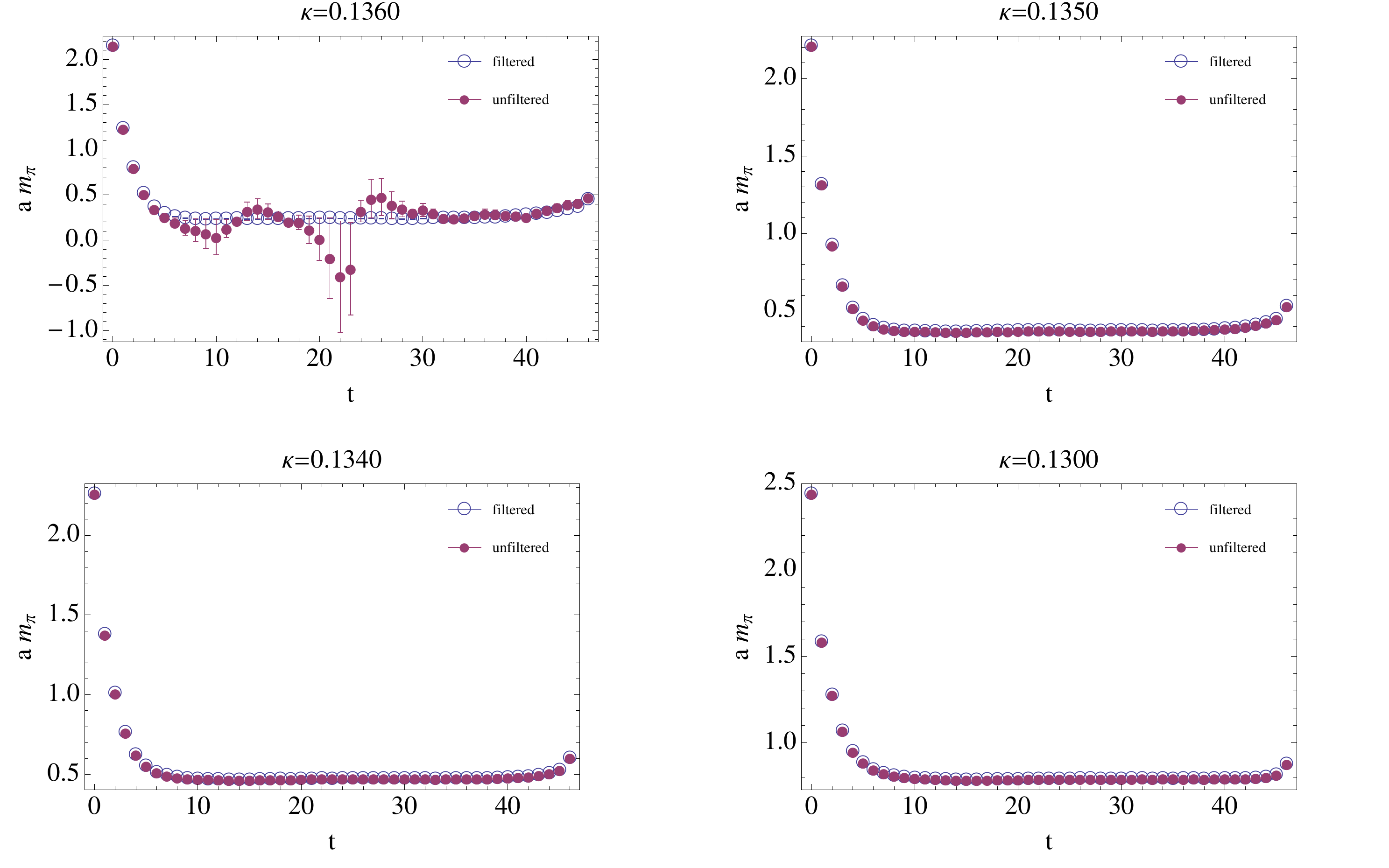} % requires the graphicx package
\caption{The pion propagator for a subset of quark masses from ensemble ${\cal E}_1$ -- 
the emty symbols are the propagator after filtering out the exceptional configurations and
the full symbols represent the original propagator.
\label{fig5}}
\end{center}
\end{figure}

One problem that we had to deal with is the presence of exceptional configurations at the lower
quark masses; the surprise was that we get exceptional configurations for pion masses as large as 
$500\MeV$ for the ensemble ${\cal E}_1$. We detected the presence of exceptional configurations
from the fact that the hadron propagators were a lot noisier on the entire ensemble than on a subset.
The method we used to detect and filter these configurations out was the following: for every propagator
of interest we looked at the smallest quark mass sector and for every time-slice we computed the mean
and its standard deviation:
\beq
\bar{G}(t) = \frac{1}{N_{\rm conf}} \sum_i G^{(i)}(t), \quad\quad \sigma_{G(t)} = 
\sqrt{\frac{1}{N_{\rm conf}} \sum_i (G^{(i)}(t)-\bar{G}(t))^2}.
\eeq
If the propagator for a configuration has at least one time-slice with \hbox{$|G^{(i)}(t)-\bar{G}(t)|>10 \sigma_{G(t)}$},
we consider that configuration to be exceptional. For a normally distributed random variable this should happen
extremely infrequently. The effect of this filtering is shown in Fig.~\ref{fig5}: we see that the heavier quark mass
propagators are unaffected while the small mass propagator is significantly improved. 

\begin{figure}[t]
\begin{center}
\includegraphics[width=11cm]{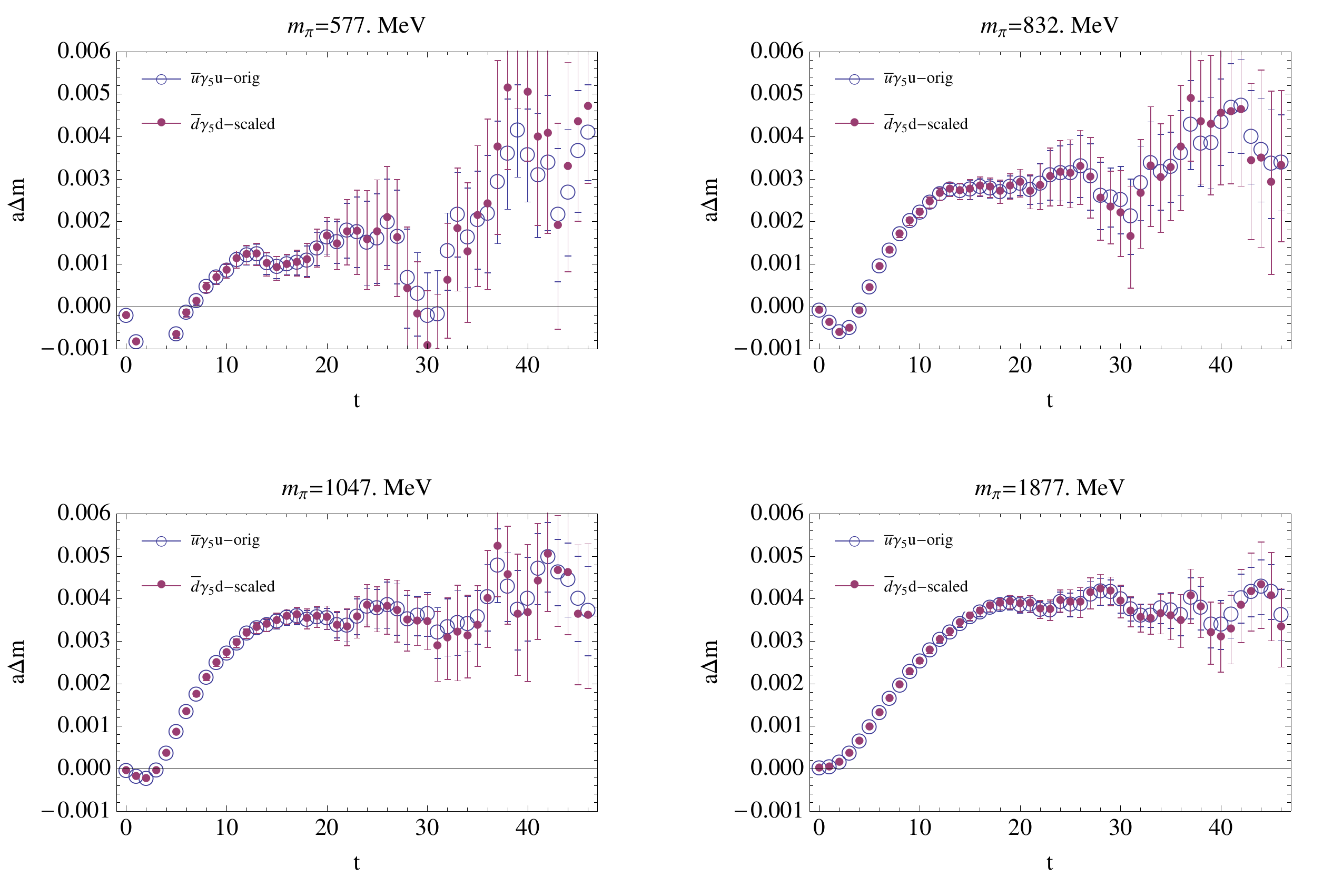} % requires the graphicx package
\caption{The effective mass shift for $\bar{u}\gamma_5 u$, empty symbols, and
scaled $\bar{d}\gamma_5 d$, full symbols.
\label{fig6}}
\end{center}
\end{figure}

The last topic we discuss is the electric field strength. In dimensionless units the value of the
field is $\eta = a^2 q E = 0.00576$. In the macroscopic realm, this is a very strong field -- on the other
hand it is comparable with the electric field a quark feels when placed about $1\fm$ away
from another quark. Thus the shift induced by this field should be comparable with the 
electromagnetic mass shift in hadrons which is known to be small. Even better, we can determine
whether the electric field is small enough by checking the scaling of the mass shift with the
increase of the electric field. We have not run the simulations with different electric field, since 
we have already seen in our previous studies that this field is small enough~\cite{Christensen:2004ca, Alexandru:2008sj}.
However, it is actually possible to check using the data that we have whether the scaling is
obeyed; since the up and down quark have the same mass in the isosipin limit but different
charges we can ask whether the mass shift for the $\bar{u}\gamma_5 u$ and $\bar{d}\gamma_5 d$
propagators scales appropriately. In Fig.~\ref{fig6} we plot the effective mass shift for these
two propagators; we scale the $\bar{d}\gamma_5 d$ propagator by a factor of $4$ (in the 
weak field limit the mass shift is proportional to the square of the charges).  It is easy to see that
the agreement is almost perfect for all quark masses; since the agreement is so good for the
effective mass shift, any mass shift extracted from these propagators will scale perfectly. We
conclude that the field we used in our simulation is weak enough.

\section{Simulation results}

In this section we present our results for the neutral pion and the neutron. The
results presented here do not include the dynamical effects: the quarks are treated
in the quenched approximation and also, the sea quarks are not charged. We
expect that these effects do not play a very important role for the masses we are
simulating, but they will be important in the chiral limit.

\begin{figure}[t]
\begin{center}
\includegraphics[width=8cm]{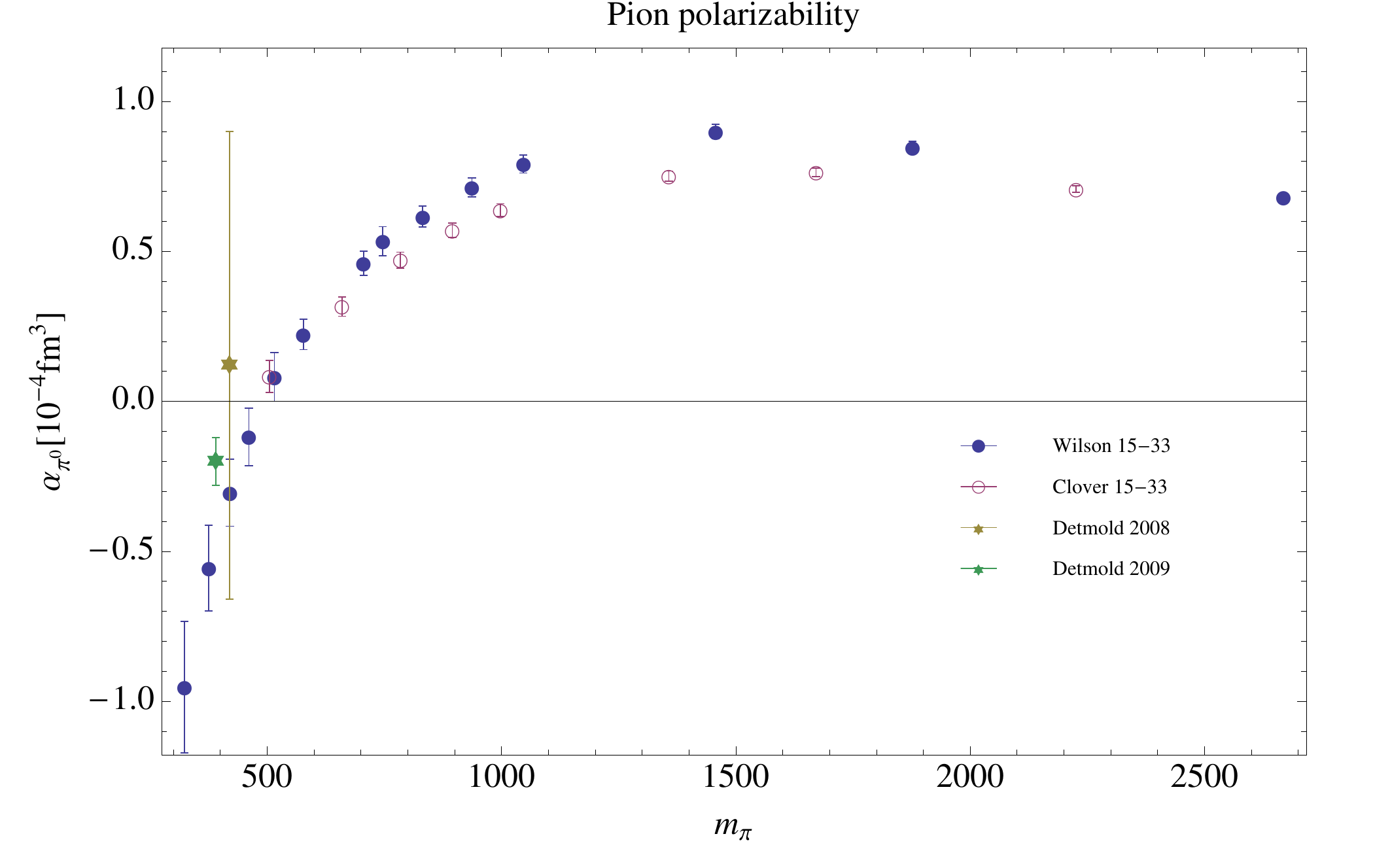} % requires the graphicx package
\caption{Pion polarizability: the blue points are our result with Wilson fermions, the empty point are the clover results and the
stars are the results from~\cite{Detmold:2008xk,Detmold:2009dx}.
\label{fig7}}
\end{center}
\end{figure}

In Fig.~\ref{fig7} we show our results for the neutral pion. We
want to stress that the results we present for the neutral pion do not include the 
disconnected piece; in the isospin limit this piece should be zero. However, the
presence of the electric field breaks the isospin symmetry even when the masses
of the up and down quarks are zero. As such, the results we present for the pion
are only an approximation. We fitted the propagators in the time window $t\in[15,33]$.
The first thing to notice is that the clover and Wilson discretizations produce similar
results: in the range $m_\pi\in[500\MeV, 800\MeV]$ they agree within error bars. 
This seems to indicate that the discretization errors are small and that we have 
a fine enough lattice spacing. For large quark masses, while qualitatively similar,
the values differ more.

It is interesting to note that for low pion masses, the polarizability seems to decrease.
In fact, around $m_\pi=500\MeV$ the polarizability turns negative and it seems to diverge.
This is in agreement with the results produced in dynamical simulations using clover
fermions~\cite{Detmold:2008xk,Detmold:2009dx}. We note that our values agree well
with the dynamical simulations, indicating that the dynamical effects are still small
at the pion masses we are studying. It should be noted that while the value for the
pion polarizability is expected to be negative, this is a result of the disconnected piece
that we have not included in our calculation. The expectation is that, for the propagator
we compute, the polarizability should be positive~\cite{Detmold:2009dx}. The
most likely culprit seems to be the finite volume effects, that are expected to be large
for polarizability -- we plan to investigate this issue further.

\begin{figure}[t]
\begin{center}
\includegraphics[width=7.5cm]{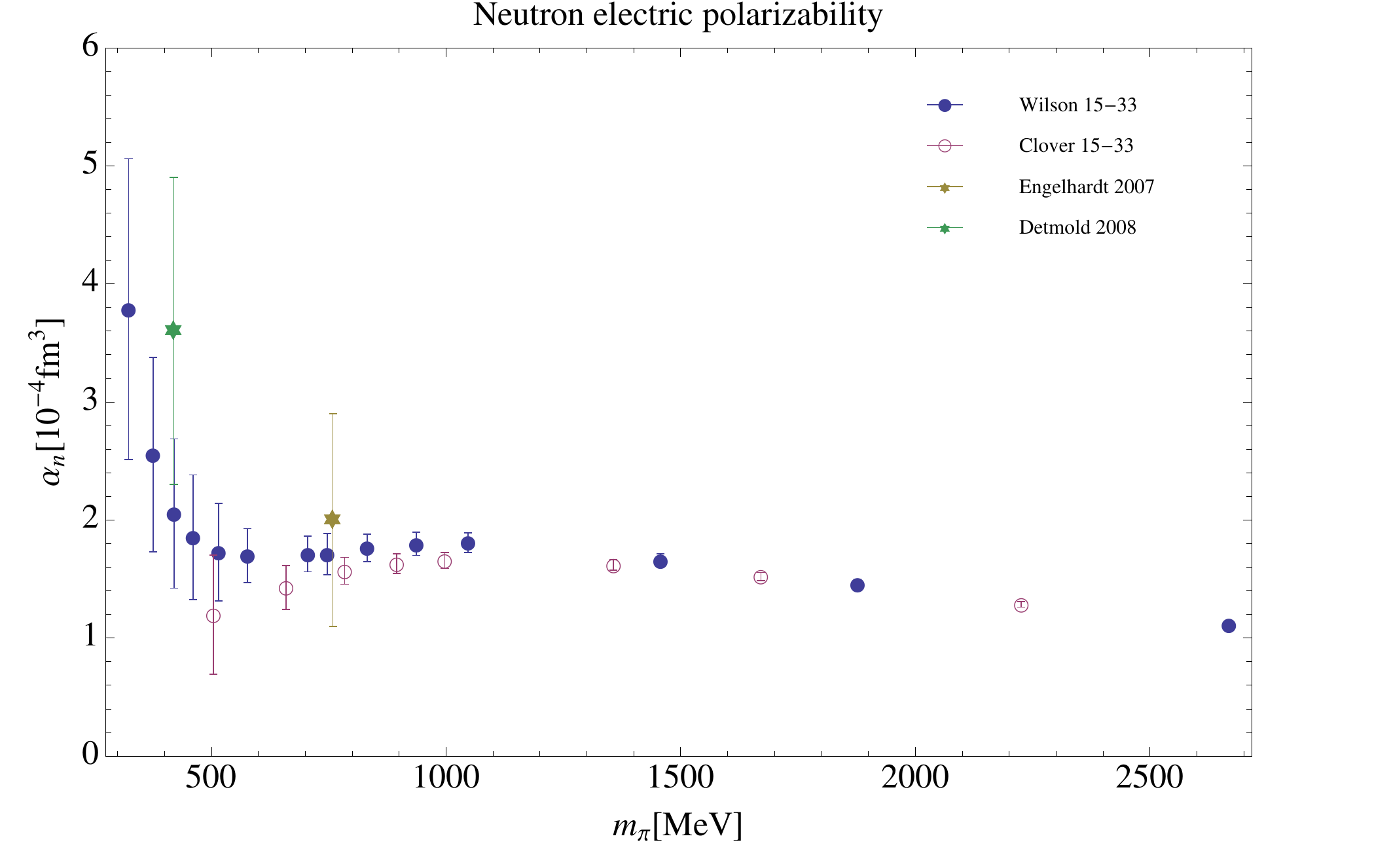} % requires the graphicx package
\includegraphics[width=7.5cm]{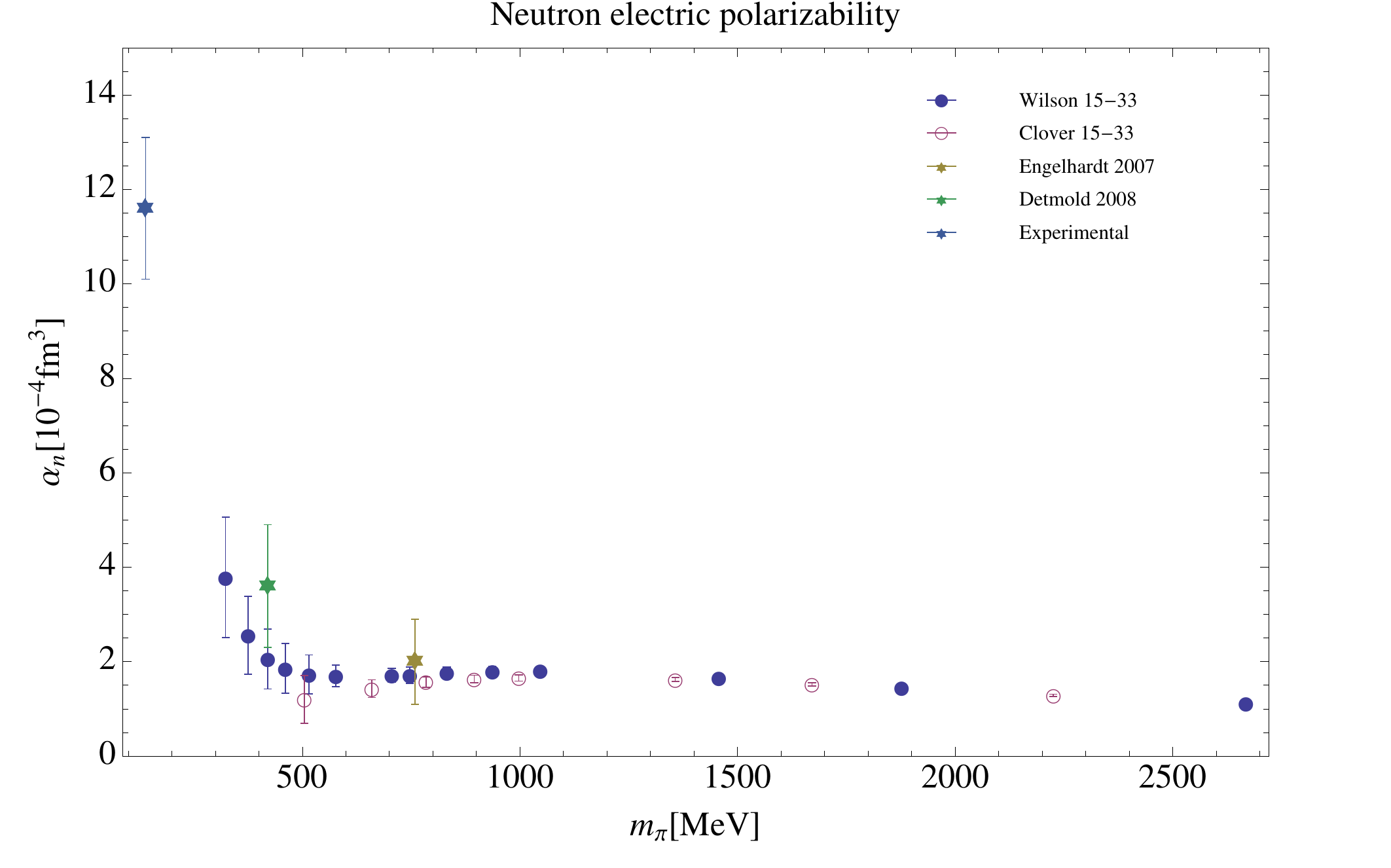} % requires the graphicx package
\caption{Neutron polarizability: the blue points are our result with Wilson fermions, the empty point are the clover results and the
stars are the results from~\cite{Detmold:2008xk,Engelhardt:2007ub} and the experimental result. The left panel doesn't
show the experimental point to allow a better view of the lattice results.
\label{fig8}}
\end{center}
\end{figure}

In Fig.~\ref{fig8} we plot our results for the electric polarizability of the neutron. Once again, we see that the Wilson
and clover results agree nicely. The results also agree well with those produced by dynamical simulations -- this
again shows that the dynamical effects are relatively small at these pion masses. Finally, the most important
feature is that for pion masses smaller than $500\MeV$ the polarizability seems to increase rapidly. We are 
finally seeing the predicted increase in the polarizability as we approach the chiral limit. Moreover, looking
at the right panel in Fig.~\ref{fig8}, it is conceivable that if the trend continues we will reproduce the experimental
result when we reach the physical limit. However, our result is still only about $3.8(1.3)\times 10^{-4}\fm^3$ for
$m_\pi = 320\MeV$, about a third of the experimental value -- there is still quite a bit to go to reach the physical limit.

To sum up, we find that the clover and Wilson fermions produce similar results for both the pion and
the neutron; this tells us that the discretization errors should be smaller than our current error bars. 
Our results are in good agreement with polarizabilities measured on dynamical ensembles which
indicates that the dynamical effects are not very large. The pion polarizability seems to be more and
more negative as we lower the quark mass, a result that is puzzling. The neutron polarizability shoots up
when the pion mass is lowered below $500\MeV$ -- this gives us hope that we can use chiral perturbation
theory to extrapolate to the physical point.

\section{Outlook}

The most important conclusion of our study is that the behavior predicted by chiral perturbation theory
for the neutron polarizability seems to appear for pion masses in the range $300\MeV$ to $500\MeV$. 
The error bars in the current study are too large to allow us to extrapolate the result, but this is mainly
due to our rather small statistics for ensemble ${\cal E}_3$. We are currently collecting more statistics
and we plan to fit the data and extrapolate it to the physical point. We also plan to run simulations at 
even lower quark masses to improve the quality of our extrapolation. The exceptional configurations
might be a problem -- in that case we plan to use a better discretization for the fermions. As a cheap
alternative, we can use nHYP fermions or, if we have to, overlap fermions since they don't suffer from
this issue.

The study produced also a puzzle regarding the polarizability of the neutral pion. As it was pointed 
out~\cite{Detmold:2009dx}, the most likely culprit is the finite volume effects. In our case the finite volume
effects are mainly due to our choice of boundary conditions. We plan to investigate the interplay between
the boundary conditions and the finite volume effects on the polarizability.

In the long run, we plan to include the dynamical effects of the sea quarks on the polarizability.
At this point, we feel that the best use for our computational resources is to pin down as best as we
can the method to determine the polarizability in the context of quenched approximation and to determine
its value at the physical point. 
The strategy that we plan to use to charge the sea quarks is reweighting and we expect that this would be very resource intensive.
To complete the calculation, we need to also use dynamically generated configurations; while this is
an important step to finalize the calculation, we feel that its contribution is the smallest while the
computational resources needed are significant. As such, this will be the last step in our program.

\end{document}